\documentclass[aps,prb,twocolumn,showpacs,groupedaddress]{revtex4}
\usepackage{graphicx}
\usepackage{bm}
\begin{document}
\title{
 Orbital ordering in undoped manganites via a generalized Peierls instability}
\author{S. Yarlagadda,$^{1,2,3}$ P. B. Littlewood,$^3$ M. Mitra,$^2$
R. K. Monu$^1$ }
\affiliation{$^1$CAMCS, Saha Institute of Nuclear Physics, Calcutta, India\\
$^2$TCMP Div., Saha Institute of Nuclear Physics, Calcutta, India\\
$^3$Cavendish Lab, Univ. of Cambridge, UK}
\date{\today}
\begin{abstract}
We study the ground state orbital ordering of $LaMnO_3$,
at weak electron-phonon coupling,
when the spin state is A-type antiferromagnet.
We determine the orbital ordering by extending
 to our Jahn-Teller system
a recently developed Peierls instability framework for
the Holstein model \cite{sdys}.
By using two-dimensional dynamic response functions corresponding to a
mixed Jahn-Teller mode, we establish that the $Q_2$ mode
determines the orbital order.
\end{abstract}

\maketitle

\section{INTRODUCTION}
Undoped manganites like $LaMnO_3$ are the parent systems for the colossal
magnetoresistive materials. It is well known that orbital ordering occurs
around 780 K resulting in a C-type orbital structure with two kinds
of orbitals alternating on adjacent sites in the $xy$ plane while
like orbitals are stacked in the $z$ direction
\cite{murakami}.
 As the temperature is further
lowered to 140 K, an A-type spin antiferromagnetic order sets in wherein
the spins are ferromagnetically aligned in the $xy$ plane with the spin
coupling in the $z$ direction being antiferromagnetic
\cite{wollan}.
To explain the
observed order several studies have been reported.
 These studies fall
into two broad classes based on the dominant cause
for the observed order.
 One class corresponds to
electron-electron (Coulombic) interaction
\cite{khomskii,ishihara,sheng,maekawa}
being the main cause
while the other class treats the cooperative
 Jahn-Teller (JT) interaction
\cite{millis,dagotto,allen,satpathy}
as the more important one. Lin and Millis \cite{lin} have made a quantitative
analysis of the effects of both interactions, generally concluding that
both pieces of physics are important, but with many subtleties.
There is further controversy about the strength of
the electron-phonon interaction with extended X-ray absorption fine structure
\cite{bianc} and pulsed neutron diffraction \cite{louca} measurements
pointing to strong interaction while some electron microscopy measurements
 \cite{mathur1,mathur2} have inferred weak coupling in the charge 
ordered phases. In that
 regime optical measurements often infer small electronic gaps\cite{optics}
 and measurements of nonlinear transport\cite{cox} have been interpreted
 as due to sliding motion of a density wave.
 
 Without addressing {\em ab initio} the issues of the quantitative strength
 of the interactions it is worth understanding how in principle a weak
 coupling theory might possibly work. The notion of JT is a molecular one,
 and the linear splitting of levels by a local distortion a useful principle
 only if the induced gap is much larger than the bandwidth (which it is not).
 Nonetheless, oxides are generally viewed as a template for strong interaction
 physics, both of the electronic and phononic variety. In this paper, we step
 back from the complexities of the full many-body theories to point out that
 the canonical model for $LaMnO_3$ has a {\em weak-coupling} generalized
 Peierls instability that reproduces qualitatively the ordering observed.
 One advantage of the simplification introduced by our approach is that we
 can study effects of adiabaticity
 that turn out to enter
 {\em logarithmically} in the ratio of electronic bandwidth to phonon frequency.

Our observation follows straightforwardly from assuming  A-type
 antiferromagnetic ordering. 
On account of strong Hund's coupling,
the transport is restricted to spin polarized electrons
 in two dimensions only, where furthermore the bands are strongly nested.
The proximity to a nesting instability allows us to employ
 the  {\it weak-coupling} framework developed
 earlier \cite{sudhakar,sdys}
and analyze
the orbital ordering by using a generalized Peierls instability approach.
However, as compared to the one-dimensional Peierls charge density wave (CDW)
approach, our higher dimensional orbital density wave (ODW) analysis is
more complicated on account of there being
two $e_g$ orbitals (with inter-orbital hopping) and two response functions
corresponding to the JT $Q_2$ and $Q_3$ distortions.
The consequences of a nesting instability on the orbital ordering
in $LaMnO_3$ were first discussed by
Yarlagadda and Mitra \cite{yarlagadda}
and later qualitatively by Efremov and Khomskii
\cite{efremov}. 

In this paper, we study the Peierls instability condition by extending
the  recently developed reliable condition involving the dynamic susceptibility
\cite{sdys}
to a mixed JT mode.
We find that $Q_2$ Jahn-Teller
distortion, as observed experimentally,
preempts other JT normal mode distortions at all
values of adiabaticity and temperature.
Furthermore, the condition of instability
(i.e., functional dependence of critical coupling on adiabaticity)
 is qualitatively similar
to that of the one-dimensional single-orbital Holstein model.
Lastly, we also find that
 mean-field approximation (in spite of being crude)
 and static Peierls instability condition (albeit erroneous)
 indicate that
$Q_2$ mode rather than $Q_3$ mode determines the orbital order.

\section{MODEL HAMILTONIAN}
\begin{figure}
\includegraphics[width=3.0in]{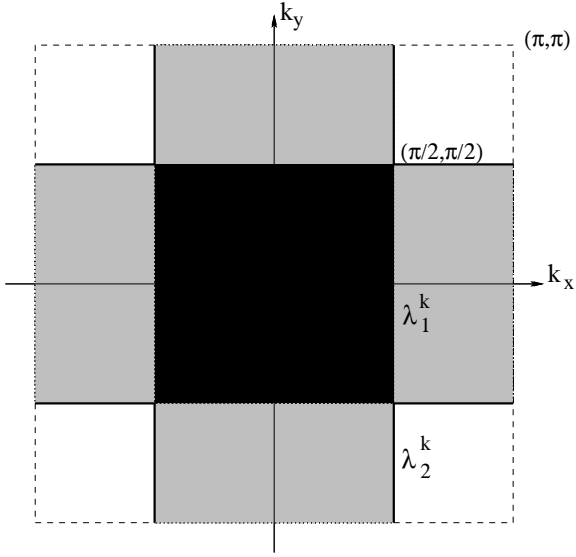}
\noindent\caption[]
{Fermi seas corresponding to the eigenenergies
$\lambda_{1,2}^{\vec{k}}$.}
\label{fig1}
\end{figure}

We will now consider manganite systems with two $e_g$ orbitals per
site and ignore spin. The Hamiltonian consists of
the kinetic term, the ionic term, and the electron-ion interaction term.
The kinetic term in momentum space is given by
\begin{equation}
H_1 = \sum _{\vec{p}}{\bf B}^{\dagger}_{\vec{p}}\cdot{\bf T}
\cdot {\bf B}_{\vec{p}} ,
\end{equation}
where $ {\bf B}^{\dagger}_{\vec{p}} \equiv
(b^{\dagger}_{1\vec p} , b^{\dagger}_{2 \vec p})$ with
$b_1$  and $b_2$
corresponding to the destruction operators
for electrons with the orthonormal wavefunctions
 $\psi_{x^2-y^2}$ and $\psi_{3z^2 -r^2}$ respectively.
Furthermore, ${\bf T}$ is a hermitian matrix with
 ${\bf T}_{1,1}=-1.5 t[\cos p_x + \cos p_y]$,
${\bf T}_{2,2}=-0.5 t[\cos p_x + \cos p_y]$, and
 ${\bf T}_{1,2}=0.5 \sqrt{3} t[\cos p_x - \cos p_y ]$.
The eigenvalues of the kinetic energy are given by
$\lambda_{n}^{\vec{p}} = -t [ \cos p_x + \cos p_y + (-1)^n \sqrt{\cos^2 p_x
+ \cos^2 p_y  - \cos p_x  \cos p_y } ]$ with $n=1,2$.
The Fermi sea corresponding to the lower eigenenergy value
$\lambda_{2}^{\vec{k}}$ is given by the union of the region
$-\pi/2 \le  k_x \le \pi/2$ (with all values of $k_y$ allowed) and
the region
$-\pi/2 \le  k_y \le \pi/2$ (with all values of $k_x$ allowed) as
shown by the shaded region (both dark and light) in Fig. \ref{fig1}.
Whereas the Fermi sea corresponding to the higher eigenenergy value
$\lambda_{1}^{\vec{k}}$
is given by the intersection of the region
$-\pi/2 \le  k_x \le \pi/2$ and the region
$-\pi/2 \le  k_y \le \pi/2$, i.e., only the dark
shaded region in Fig. \ref{fig1}.
 Since the number of electrons is equal to
the number of sites, the total area
occupied by both Fermi seas is equal to the area of the
Brillouin zone ($4 \pi^2$). Furthermore, the Fermi surface corresponds
to $\lambda_{n}^{\vec{k}} = 0$.

 The electron-phonon interaction term is  given by
\begin{eqnarray}
H_3 = g \omega_0 \sqrt{2 M \omega_0} \sum_{j} [ &&
\!\!\!\!\!\!
Q_{2j}
(b^{\dagger}_{1j} b_{2j} + b^{\dagger}_{2j} b_{1j})
\nonumber \\
&&
\!\!\!\!\!\!
 +
Q_{3j}
(b^{\dagger}_{1j} b_{1j} - b^{\dagger}_{2j} b_{2j}) ] ,
\end{eqnarray}
while the phononic part of the Hamiltonian is given by
\begin{equation}
H_{2} = \omega_0
\sum_{j}
\sum_{l=2,3} f^{\dagger}_{lj} f_{lj} ,
\end{equation}
where $f_{lj} + f^{ \dagger}_{lj} = \sqrt{2 M \omega_0} Q_{lj}$.
Since we are interested in understanding orbital order, our Hamiltonian
 does not contain breathing mode distortions.

\section{PEIERLS INSTABILITY}
In this section, to understand orbital ordering at weak electron-phonon
coupling,
we consider the Peierls instability condition
by using the dynamic susceptibility
instead of the static one (see Appendix A for a justification).
The cooperative Jahn-Teller effect requires  compatible
distortions on adjacent sites which implies that
the ordering wavevector in two-dimensions is given by $\vec{Q}
\equiv (\pi , \pi )$.
We expand the free energy to quadratic order in the relevant
degree of freedom (i.e.,  density $n$ of electrons in an appropriate occupied
 orbital) as follows:
\begin{eqnarray}
F = \sum_{\vec{q} = \pm \vec{Q}}
 &&
\left [ -
 \frac{n_{\vec{q}} n_{-\vec{q}}}
{2 {\rm Re}\chi_{\phi}(\vec{q}, \omega)}
 + g \omega_0 n_{-\vec{q}} [\langle
f^{\dagger}_{\phi -\vec{q} } \rangle
 + \langle f_{\phi \vec{q}} \rangle ] \right .
\nonumber \\
&&
\left .
 +\omega_0 \langle f^{\dagger}_{\phi \vec{q}} \rangle
 \langle f_{\phi \vec{q}} \rangle \right ] ,
\end{eqnarray}
where $f_{\phi j} + f^{\dagger}_{\phi j} = \sqrt{2 M \omega_0} Q_{\phi j}$.
Here
 $Q_{\phi}$ is the dominant mode defined as
$Q_{\phi} \equiv Q_{3} \cos(2 \phi) + Q_2 \sin(2 \phi)$
where only orbitals $\psi_{x^2-y^2} \cos(\phi) + \psi_{z^2} \sin(\phi)$
or their orthonormal orbital states
 $-\psi_{x^2-y^2} \sin(\phi) + \psi_{z^2} \cos(\phi)$
are occupied.
The order parameter corresponding to phonons is given by
$\langle f_{\phi \vec{Q}} \rangle =
 | \langle f_{\phi \vec{Q}} \rangle |e^{i\Theta} $.
Using reflection symmetry, we first note that
$\chi_{\phi}(\vec{Q}, \omega )
= \chi_{\phi}(-\vec{Q}, \omega )$,
$n_{\vec{Q}} = n_{-\vec{Q}}$,
$\langle f_{\phi \vec{Q}} \rangle  =\langle f_{\phi -\vec{Q}} \rangle$,
and $\langle f^{\dagger}_{\phi \vec{Q}}\rangle = 
\langle f^{\dagger}_{\phi -\vec{Q}} \rangle$.
For $g > 0$, free energy minimum occurs at $\Theta = \pi$.
Minimizing $F$, with respect to
$|\langle f_{\phi \vec{Q}} \rangle |$,
 yields
$|\langle f_{\phi \vec{Q}} \rangle| =  g n_{\vec{Q}}$.
Thus, we get
\begin{eqnarray}
F = -
 \left[ \frac{1+ 2 g^2 \omega_0 {\rm Re} \chi_{\phi}(\vec{Q}, \omega)}
{{\rm Re}\chi_{\phi}(\vec{Q}, \omega )} \right ] n_{\vec{Q}} n_{-\vec{Q}} .
\end{eqnarray}
On defining the effective susceptibility as
\begin{eqnarray}
\chi^{eff}_{\phi } \equiv
\frac{{\rm Re}\chi_{\phi}}{1+ 2 g^2 \omega_0 {\rm Re}\chi_{\phi}} ,
\end{eqnarray}
the Peierls instability condition is given
 by
\begin{eqnarray}
1+2 g^2 \omega_0 {\rm Re} \chi_{\phi} (\vec{Q}, \omega_0) =0 ,
\label{peierls}
\end{eqnarray}
and leads to the divergence of
$\chi^{eff}_{\phi}(\vec{Q}, \omega_0)$.
We take
 $\omega = \omega_0$
 in
$\chi^{eff}_{\phi}(\vec{Q}, \omega)$
 because $\omega_0$ is the natural frequency for
lattice distortion. A better explanation for choosing $\omega = \omega_0$
is given in Appendix A.

We need to determine at what value of $\phi$ one gets the
largest value of
${\rm Re} \chi_{\phi} (\vec{Q}, \omega_0)$.
Then one can determine which
normal mode gives the lowest value of $g=g_c$
satisfying the Peierls instability condition.
 Note that, as the rotational angle
$\phi$ (for $0 \le \phi \le \pi/2$) is varied, all possible
normal modes are spanned starting from $Q_3$ at $\phi=0$
to $Q_2$ at $\phi = \pi/4$ and then to $-Q_3$ at $\phi = \pi/2$.
 The dynamic susceptibility is given by
\begin{eqnarray}
 \chi_{\phi} (\vec{q}, \omega) = \sum_{n}
\left [
 \frac{|\langle n | \rho Q_{\phi} (\vec{q}) |0 \rangle |^2 }
{\omega - \xi_{n0} + i \eta}
 - \frac{|\langle 0 | \rho Q_{\phi} (\vec{q}) |n \rangle |^2 }
{\omega + \xi_{n0} + i \eta}
\right ] ,
\end{eqnarray}
where
\begin{eqnarray}
 \rho Q_{\phi} (\vec{q}) \equiv
\sum_{\vec{k}}
&&
\left [
b^{\dagger}_{1 \vec{k}+\vec{q}}b_{1 \vec{k}}
- b^{\dagger}_{2 \vec{k}+\vec{q}}b_{2 \vec{k}}
\right ] \cos (2 \phi )
\nonumber \\
&&
+
\left [
b^{\dagger}_{1 \vec{k}+\vec{q}}b_{2 \vec{k}}
+ b^{\dagger}_{2 \vec{k}+\vec{q}}b_{1 \vec{k}}
\right ] \sin (2 \phi ) .
\end{eqnarray}

Then, after some algebra, one gets
\begin{eqnarray}
 \!\!\!\!\!
 {\rm Re} \chi_{\phi} (\vec{q}, \omega_0) =
 \sum_{\vec{k}, \alpha, \beta}
&&
 \!\!\!\!\!
\left [
 \frac
{
\langle c^{\alpha \dagger}_{\vec{k}} c^{\alpha}_{\vec{k}}\rangle
-
\langle c^{\beta\dagger}_{\vec{k}+\vec{q}} c^{\beta}_{\vec{k}+\vec{q}}\rangle
}
{\omega_0 +  \lambda^{\vec{k}}_{\alpha} - \lambda^{\vec{k}+\vec{q}}_{\beta}}
\right ] \times
\nonumber \\
&&
 \!\!\!\!\! \!\!\!\!\!\!\!\!\!\!\!\!
\cos^2 \left [
\frac{\theta_{\vec{k}+\vec{q}} +\theta_{\vec{k}}
+ (\alpha+\beta)\pi
}{2} + 2\phi
\right ] ,
\label{chi-phi}
 \end{eqnarray}
where $\alpha = 1,2$; $\beta=1,2$;
$( c^{1\dagger}_{\vec{k}}, c^{2\dagger}_{\vec{k}}) = 
(b^{\dagger}_{1 \vec{k}}, b^{\dagger}_{2 \vec{k}}) \cdot {\bf M}$,
 ${\bf M}$ is the diagonalizing matrix for
the kinetic matrix ${\bf T}$ with
${\bf M}_{1,1}=\sin(\theta_{\vec{k}}/2)$,
${\bf M}_{2,2}=-\sin(\theta_{\vec{k}}/2)$, and
${\bf M}_{1,2}=\cos(\theta_{\vec{k}}/2)$.
It is interesting to note that,
for symmetric wavevectors $\vec{q} =(q,q)$,
 there is no coupling between the density
operators corresponding to $Q_2$ and $Q_3$ modes
 because the inter-orbital hopping
 ${\bf T}_{1,2}=0.5 \sqrt{3} t[\cos p_x - \cos p_y ]$
is asymmetric with respect to interchange of momenta $p_x$ and $p_y$.
 Thus for
 $\vec{q} =(q,q)$, we obtain
\begin{eqnarray}
 \!\!\!\!\!
\chi_{\phi} (\vec{q}, \omega_0) =
 \chi_{3} (\vec{q}, \omega_0) \cos^2 (2\phi ) +
 \chi_{2} (\vec{q}, \omega_0) \sin^2 (2\phi ) ,
\label{chi-decouple}
 \end{eqnarray}
where $\chi_{2,3}$ correspond to JT modes $Q_{2,3}$

\subsection{Static Instability Case}
\begin{figure}[t]
\includegraphics[width=3.0in]{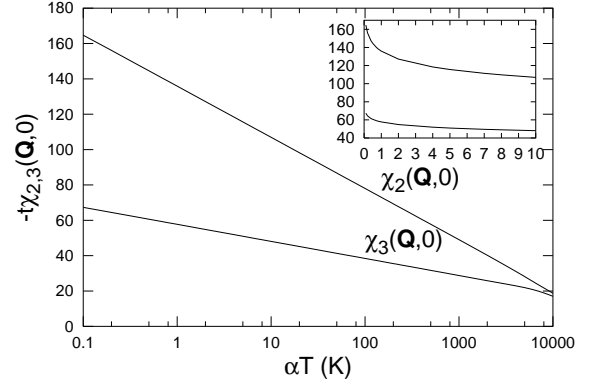}
\noindent\caption[]
{Plot of $\chi_{2,3} (\vec{Q},0)$ as a function of
scaled temperature $\alpha T$ at $\alpha t=1.0 ~ eV$ and
$ {\bf Q}\equiv \vec{Q} = (\pi,\pi)$.}
\label{fig2}
\end{figure}

Now, although the static Peierls instability condition
$1+2 g^2 \omega_0 \chi_{\phi} (\vec{Q},0) =0 $
erroneously predicts instability even for vanishing values of $g$,
it can still help identify which normal mode produces the Jahn-Teller
instability. We will first present results for the static susceptibilities
$\chi_{2,3} (\vec{Q},0)$.
From the plot of $\chi_{2,3} (\vec{Q},0)$
(shown in  Fig. \ref{fig2}) as a function of
scaled temperature $\alpha T$
(with $\alpha$ being a scaling parameter and
 hopping term $\alpha t$ set equal to 1.0 eV) we see
 that they diverge logarithmically as $T \rightarrow 0$
with $\chi_2$ diverging faster than $\chi_3$.
At 0 K, both $\chi_2 (\vec{Q} , 0)$ and
$\chi_3 (\vec{Q}, 0)$ produce a divergence because of the fact that
$ \lambda^{\vec{k}+\vec{Q}}_{1} = -\lambda^{\vec{k}}_{2}$ and that
the Fermi energy is zero. Furthermore, the ratio
 $\chi_2 (\vec{Q} , 0)/ \chi_3 (\vec{Q} , 0)=3$
at 0 K (see Appendix B for details).
As can be seen from Fig. \ref{fig2},
$\chi_{2,3} (\vec{Q}, 0)$ vary logarithmically with $k_B T/t$
for $t/k_B T > 2$ and thus have the form
\begin{eqnarray}
{\rm Re}[-t \chi_{2,3} (\vec{Q}, 0)] = m_{2,3} \ln (t /k_B T )
 + \kappa_{2,3} .
\label{chiT}
 \end{eqnarray}
We find that $m_{2} \approx 12.6$ ($m_3 \approx 4.2$)
and $\kappa_2 \approx 18.3$ ($\kappa_3 \approx 18.5$)
with the ratio $m_2/m_3$ taking the expected value $3$.
 Thus it appears that
$Q_2$ mode is likely to dictate the orbital ordering.

\begin{figure}[b]
\includegraphics[width=3.0in]{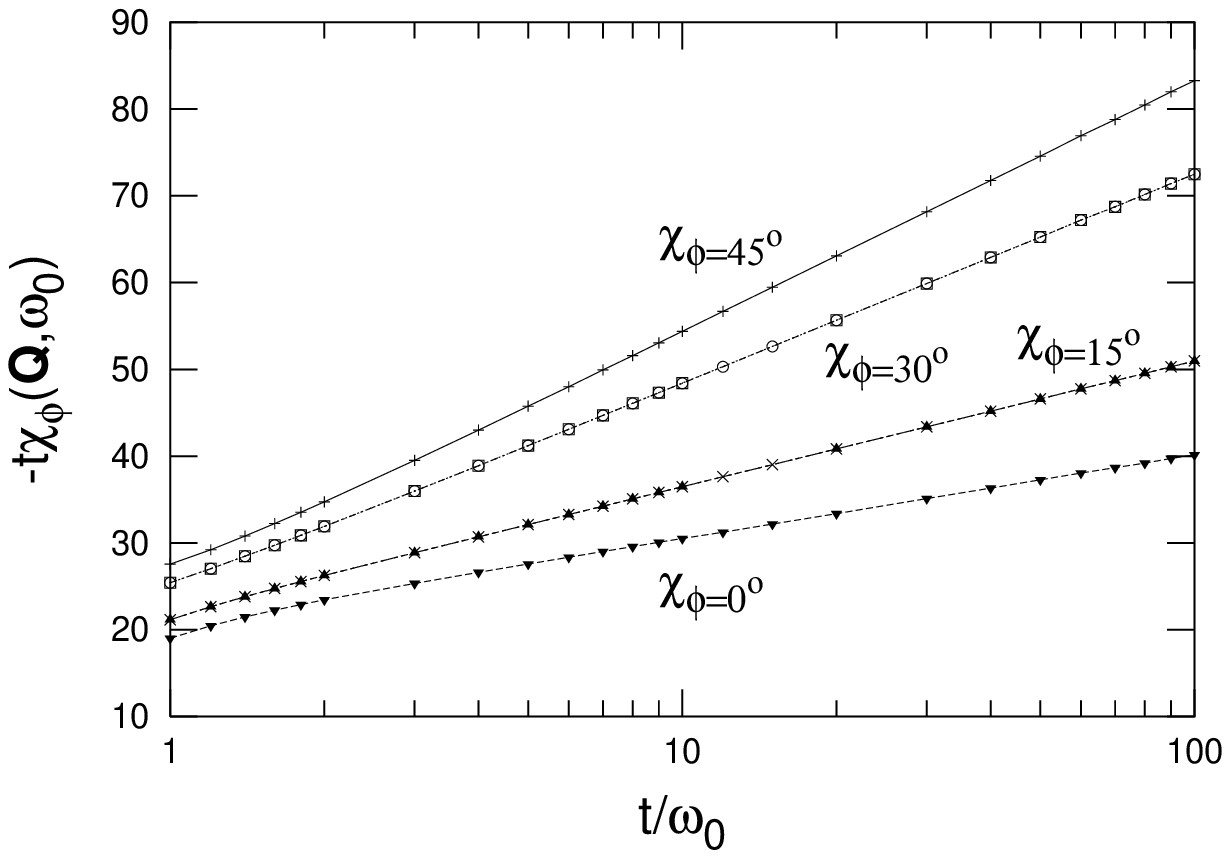}
\noindent\caption[]
{Plot of ${\rm Re} \chi_{\phi}$ as a function of the adiabaticity $t/\omega_0$
for various values of $\phi$. $\phi = 0^{\circ}$ ($45^{\circ}$) corresponds
 to $\chi_3$ ($\chi_2$). }
\label{fig4}
\end{figure}

\begin{figure}[t]
\includegraphics[width=3.0in]{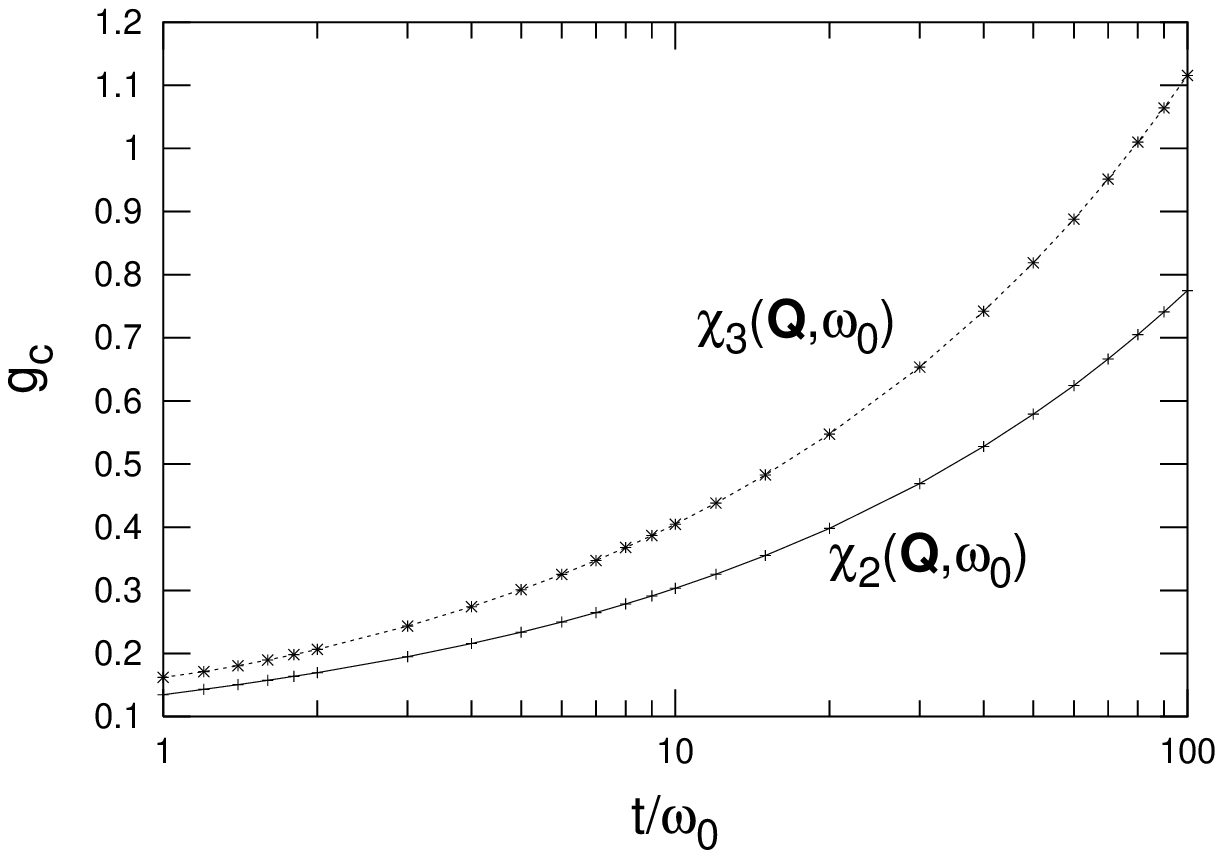}
\noindent\caption[]
{Plot of the critical coupling $g_c$ as a function of adiabaticity $t/\omega_0$
for the susceptibilities $\chi_2$ and $\chi_3$.}
\label{fig5}
\end{figure}

\begin{figure}[b]
\includegraphics[width=3.0in]{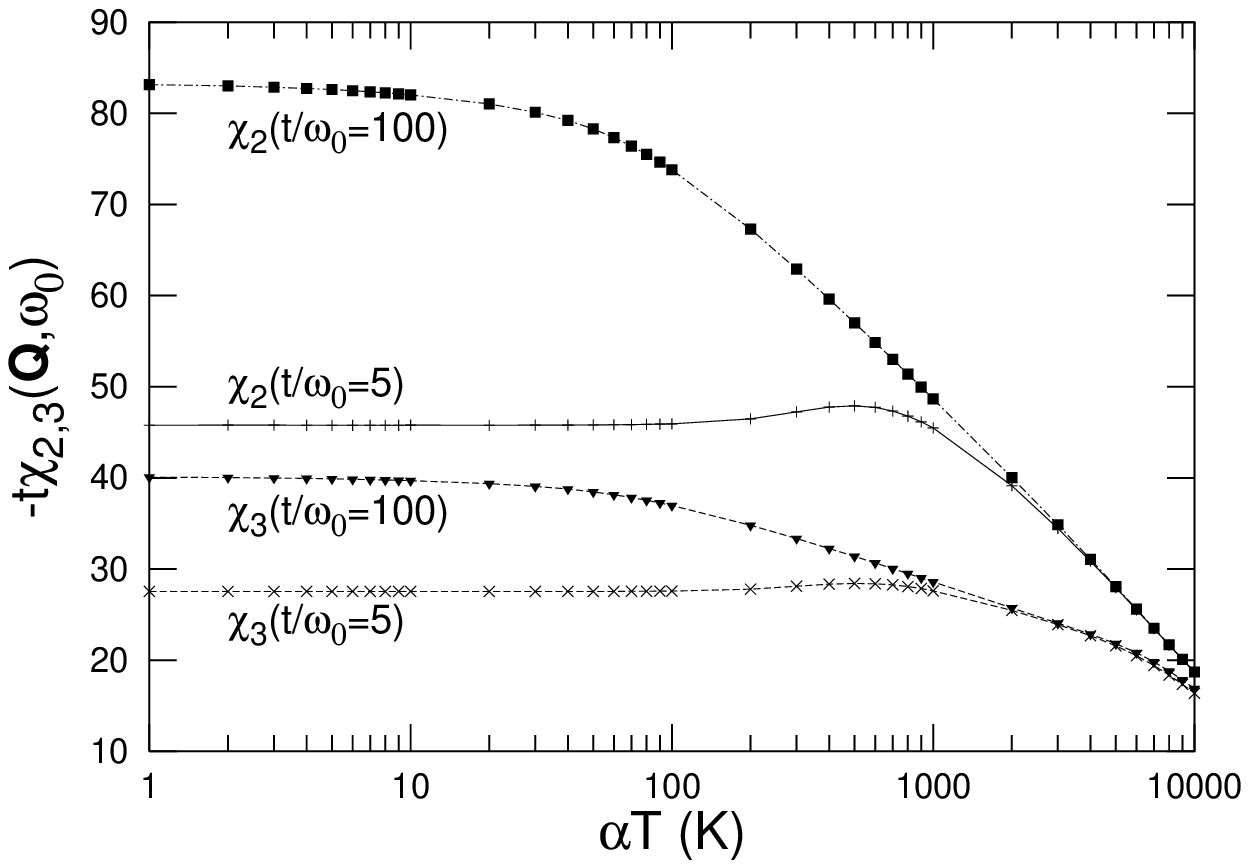}
\noindent\caption[b]
{Plot of the susceptibilities
${\rm Re} \chi_{2,3} (\vec{Q}, \omega_0)$ as a function 
of the scaled temperature
$\alpha T$ for values of scaled hopping $\alpha t = 1.0$ eV
 and adiabaticity
$t/\omega_0 =5, ~100$.}
\label{fig6}
\end{figure}

\subsection{Dynamic Instability Case}
While both the static Peierls instability condition and the
 mean-field energy analysis (see Appendix C)
 depend only on the polaron size parameter
($g^2\omega_0/t$), here for the dynamical Peierls instability
condition [of Eq. (\ref{peierls})]
there are two relevant parameters -- namely
adiabaticity parameter $t/\omega_0$ and electron-phonon coupling $g$.
We find that for any value of the adiabaticity parameter $t/\omega_0$
 the maximum value of
${\rm Re} \chi_{\phi} (\vec{Q}, \omega_0)$ occurs at $\phi=\pi/4$
which corresponds to $Q_2$ mode. In Fig. \ref{fig4}, using Eq. (\ref{chi-phi}),
a variation of ${\rm Re} \chi_{\phi} (\vec{Q}, \omega_0)$
 (at $0$ K) is plotted for a few
representative values of $\phi=0,\pi/12,\pi/6,\pi/4$.
The curves for
${\rm Re} \chi_{\pi/12} (\vec{Q}, \omega_0)$ and
 ${\rm Re} \chi_{\pi/6} (\vec{Q}, \omega_0)$ (in
Fig. \ref{fig4}) verify
 Eq. (\ref{chi-decouple}).
Furthermore, we also found numerically
that ${\rm Re} \chi_{\phi} (\vec{Q}, \omega_0)$ [given by Eq. (\ref{chi-phi})]
 is symmetric about $\phi = \pi/4$ -- a fact that follows from
Eq. (\ref{chi-decouple}).

Quite strikingly, all the
${\rm Re} \chi_{\phi} (\vec{Q}, \omega_0)$ vary logarithmically
with the adiabaticity $t/\omega_0$ for $t/\omega_0 > 2$
and have the form
\begin{eqnarray}
{\rm Re}[-t \chi_{\phi} (\vec{Q}, \omega_0)] = m_{\phi} \ln (t /\omega_0 )
 + \kappa_{\phi} .
\label{chiomega}
 \end{eqnarray}
 We find that
$m_{\pi/4} \approx 12.6$ ($m_0 \approx 4.2$)
 and $\kappa_{\pi/4} \approx 25.5$ ($\kappa_{0} \approx 20.9$).
Interestingly,  the slopes in Eq. (\ref{chiomega})
are the same as those in
Eq. (\ref{chiT}).
The ratio of the slopes $m_{\pi/4}$/$m_0$ = 3 as expected
from the fact that $\chi_2 (\vec{Q}, 0)$/$\chi_3 (\vec{Q}, 0)$ = 3 at $0$ K.
Furthermore, this logarithmic dependence is
 quite like that for the
Holstein model. Using the dynamic Peierls instability condition,
similar to the Holstein model case,
 we are lead to an instability condition
of the form
$\omega_0 = a_1 t e^{- a_2 t/g^2 \omega_0}$ where $a_{1,2}$ are constants.
 We also calculated the
critical value of the electron-phonon coupling
$g_c$ at which the instability occurs if only $Q_2$ mode
or only $Q_3$ mode is excited. We find that the value of $g_c$ increases
monotonically with the adiabaticity parameter (similar to the Holstein model)
and that, as expected, the $g_c$ value is the smallest for $Q_2$ distortion
(as can be seen in  Fig. \ref{fig5}) at any value of $t/\omega_0$.

We have also studied the temperature dependence of the dynamical
susceptibilities (as shown in Fig. \ref{fig6}) and find that at low temperatures
the curves are constant with the extant of the constant region increasing
 as $t/\omega_0$ decreases. Such a behavior is consistent
with the expectation that
${\rm Re} \chi_{\phi} (\vec{Q}, \omega_0)$ is constant
over the region $k_B T << \omega_0$. Furthermore, at higher temperatures
the susceptibilities for various adiabaticities merge. For instance,
when $\alpha T$ attains a value of around $300$ K, curves for
 $t/\omega_0 = 100$ and $\infty$ merge (as can be seen from  Figs.
\ref{fig2} and \ref{fig6}); and for $\alpha T$ around $2000$ K, curves for
$t/\omega_0 = 100$ and $5$ merge.
 The high temperature behavior too
is understandable because one expects the effect of  non-zero value
of $\omega_0$ to vanish when $k_B T >> \omega_0$.
 At the Jahn-Teller orbital ordering temperature
 of $780$ K and
for realistic values of both
 $t$ and $\omega_0$ (i.e., for 0.15 eV $\le t \le$ 0.38 eV \cite{arima}
and for
0.06 eV $\le$ $\omega_0$ $\le$ 0.07 eV \cite{iliev}),
range of the critical coupling [as obtained from
Eq. (\ref{peierls}) and Fig. \ref{fig6}]
is $0.2 \le g_c \le 0.28$. 
For instance, at $T = 780$ K, 
 $t=0.2$ eV [and hence $\alpha = 5$ in Fig. (\ref{fig6})], and
 $\omega_0  =$ 0.07 eV,
we get $g_c \approx 0.21$.
Lastly, for $k_B T >> \omega_0$ but $k_B T/t < 0.5$, the curves
display a logarithmic dependence on $k_B T /t$ which is in tune
with the logarithmic dependence on $\omega_0/t$ of the susceptibility
of the Holstein model when $\omega_0/t < 0.5$ (see Ref. \onlinecite{sdys}).

\section{CONCLUSIONS}
We will now discuss the general features of the orbital-ordering
instability and compare it with the Peierls instability in
the Holstein model.
For the Holstein model, at $0$ K,
the mean-field approximation gives a gap $\Delta$ of the form \cite{hamer1}
 \begin{eqnarray}
\Delta = 8 t e^{-\pi t/g^2 \omega_0} .
 \end{eqnarray}
In our case as well, we find that the gap is given by
 \begin{eqnarray}
\Delta_{2,3} \approx d_{2,3}^1 t e^{-d_{2,3}^2 t/g^2 \omega_0} ,
 \end{eqnarray}
where $d_{2,3}^{1,2}$ are constants and
 $\Delta_{2,3} \approx g^2 \omega_0 |c_{2,3}|$ with $c_{2,3}$ 
being amplitudes of orbital density waves defined in Appendix C.
It should however be noted that, when $\Delta/\omega_0 << 1$,
mean-field gives erroneous results. For instance, it predicts
a gap even when the electron-phonon coupling  $g$ is small.
Although, mean-field approximation is inaccurate at the transition, it
can still help us figure out which of the two JT modes is dominant.
As shown in Appendix C, mean-field correctly shows that $Q_2$ mode
prevails over $Q_3$ mode.

Next, in the Holstein model \cite{sdys,hamer2}, at 0 K and $t/\omega_0 > 2$,
the actual instability condition
is given by
 \begin{eqnarray}
\omega_0 = 8 t e^{-\pi t/g^2 \omega_0} .
 \end{eqnarray}
For our JT system too, at $k_B T << \omega_0$
($k_B T >> \omega_0$) and
when  $t/\omega_0 > 2$  ($t/ k_B T > 2$),
the instability is of the form
 \begin{eqnarray}
\omega_0 ~  (\gamma k_B T)= a_1 t e^{- a_2 t/g^2 \omega_0} ,
 \end{eqnarray}
and thus, like the Holstein model,
 has an essential singularity at $g =0$.
We also note that one cannot get the correct Peierls instability condition
 by the approximation $P_{\phi}^2/(2M) +K Q_{\phi}^2/2
\approx KQ_{\phi}^2/2$ for the normal mode distortion even when
$t/\omega_0$ is large.
 This is because,
when  $P_{\phi}^2/(2M) = 0$,
 the double commutator for
the distortion $Q_{\phi}$ becomes zero,
\begin{equation}
\ddot{Q}_{\phi \vec{Q}} = - [[Q_{\phi \vec{Q}}, H], H ] =0 ,
\label{A_tt}
 \end{equation}
which implies that 
phase transition always occurs!

In summary, we observe that owing to the one-dimensional like
Fermi surface at zero doping in manganites (as shown in Fig. 1),
there are strong similarities of the above mentioned nature
 between our JT system and the
one-dimensional Holstein model.
The one-dimensionality of our manganite system is a result of the
flatness of the Fermi surface [as can be seen, for instance,
 from Eq. (\ref{chi2chi3})].
When  $t/ [max \{\omega_0, k_B T, \Delta_{\phi} \}] > 2$,
we find that the susceptibility
${\rm Re}[-t \chi_{\phi} (\vec{Q}, \omega_0)]$
varies logarithmically with respect to
$t/ [max \{\omega_0, k_B T, \Delta_{\phi}\}]$
and has the general form
\begin{eqnarray*}
{\rm Re}[-t \chi_{\phi} (\vec{Q}, \omega_0)] = m_{\phi}
\ln (t /[max \{\omega_0, \gamma k_B T, \Delta_{\phi} \} ]) + \kappa_{\phi} ,
\label{chiTomega}
 \end{eqnarray*}
with $\gamma \approx 1.77 $ and both $m_{\phi}$ and $\kappa_{\phi}$  being
given by Eq. (\ref{chiomega}).
Using this logarithmic relation and the generalized Peierls instability
 condition of Eq. (\ref{peierls}), one obtains the explicit form
 of the instability condition.

In conclusion, we have studied orbital ordering for the ground state
of the undoped manganite systems in the weak electron-phonon
coupling regime $g \omega_0 /t < 1$.
We employ the generalized dynamic Peierls instability condition
$1+2 g^2 \omega_0 {\rm Re} \chi_{\phi} (\vec{Q}, \omega_0) =0 $
to figure out which normal mode or combination
of normal modes causes the instability.
It is also important to note that the dynamic
 Peierls instability condition does not suffer from the
problem of predicting CDW instability at vanishingly small
electron-phonon coupling (i.e., $g \rightarrow 0$)
as does the usual static Peierls instability condition
[$1+2 g^2 \omega_0 \chi_{\phi} (\vec{Q}, 0) =0 $].
We find that $Q_2$ Jahn-Teller
distortion produces the first instability and thus preempts other
normal mode distortions.
Thus the two-dimensional
orbital ordering, in the ferromagnetic planes of the observed
A-type antiferromagnetic
state, is governed by
the $Q_2$ JT mode
being cooperatively excited in the system. Hence, we
find that the experimentally
observed order can be explained even without considering electron-electron
interactions.

Before we close, a few general discussions are in order.
Above the magnetic transition temperature $T_N$,
where orbital structure does not
change much, transport is permitted in the
third direction and the Fermi surface for three dimensions should be
considered. Then,
  although the bands are not flat, we still have the nesting condition
  $\lambda_1^{\vec{k}+\vec{Q}} = -\lambda_2^{\vec{k}}$ for $\vec{Q}
 = ( \pi,\pi,\pi ) $ and hence the static susceptibilities will diverge.
 However, the experimental ordering wavevector is $(\pi , \pi, 0 )$
and not $(\pi,\pi,\pi)$.
To get the observed ordering one will have to incorporate additional
physics such as octahedral tilting.
Next, at non-zero temperatures below $T_N$,
hopping in the third direction is small but non-zero owing
to the non-saturation in A-type antiferromagnetic order.
Then flatness (one-dimensionality)
of the Fermi surface would be lost. However, hopping in the third direction
increases with temperature and the situation is different from that mentioned
in Ref. \onlinecite{wei}
where, since the hopping in the transverse direction decreases with
increasing temperature, re-entrant behavior could occur.
Lastly,
electron-electron interactions can have an effect on the nesting
conditions as pointed out by Kugel, Sboychakov, and Khomskii \cite{kugel}.
These authors find that electron-electron interactions lead to the
occurrence of nesting at a density of less than an electron per site.
However, in this work, Luttinger's theorem is violated 
[see Fig. 5(e) in Ref. \onlinecite{kugel}] and implications 
of that should be investigated for non-Fermi liquid behavior.
If, indeed in a full-fledged calculation, beyond the
Hubbard I approximation, nesting (with flat Fermi surface) occurs at a
lower density, then a corresponding ODW
instability condition should be re-analyzed for such a situation.

\section{ACKNOWLEDGEMENTS}
One of the authors (S.Y.) would like to thank S. Datta and R. Ramakumar
for useful discussions. This work was partially funded
by UKIERI and CAMCS of SINP.

\appendix
\section{}
We shall give a heuristic justification for the use of dynamic
 susceptibility
in the Peierls instability condition
\begin{eqnarray}
1+2 g^2 \omega_0 {\rm Re} \chi_{\phi} (\vec{Q}, \omega_0) =0 .
\label{peierlsA}
\end{eqnarray}
Let $Q_{\phi \vec{Q}}$ be the dominant normal mode
distortion operator (at wavevector $\vec{Q}$)
 in the Fourier transformed space.
 We know that the double time derivative
of the operator $Q_{\phi \vec{Q}}$ is given by
 \begin{equation}
\ddot{Q}_{\phi \vec{Q}} = - [[Q_{\phi \vec{Q}}, H], H ] .
\label{Q_tt}
 \end{equation}
Then on taking matrix elements we get
\begin{equation}
\langle \Phi_{1} | \ddot{Q}_{\phi \vec{Q}} |\Phi_0 \rangle  = -
(E_{\Phi_{1}} - E_{\Phi_0})^2 \langle \Phi_1 | Q_{\phi \vec{Q}}
 |\Phi_0 \rangle ,
\label{X_matrix}
\end{equation}
where $\Phi_n$ is an eigenstate with $n$ phonons all of which are in the state
$\vec{Q}$.
 When $\omega_{eff}^2 \equiv (E_{\Phi_1} - E_{\Phi_0})^2 \le 0$,
instability occurs for transition from $|\Phi_0 \rangle $ to
$|\Phi_{1} \rangle$   provided that
$\langle \Phi_{1}  | Q_{\phi \vec{Q}} |\Phi_0 \rangle \neq 0  $.
Now, at weak electron-phonon couplings  (i.e., when $g\omega_0/t < 1$)
\begin{eqnarray}
E_{\Phi_1} - E_{\Phi_0} && = \omega_0 + {\rm Re}\Sigma_{\phi}
 (\vec{Q},\omega_0)
\nonumber
\\
&& = \omega_0
+ g^2 \omega_0^2 {\rm Re} \chi_{\phi} (\vec{Q}, \omega_0) ,
\end{eqnarray}
where $\Sigma_{\phi}$
 is the self-energy corresponding to mode $Q_{\phi \vec{Q}}$. Thus, when
\begin{eqnarray}
\omega_{eff}^2 = \omega_0^2[
1+2 g^2 \omega_0 {\rm Re} \chi_{ \phi} (\vec{Q}, \omega_0)] =0 ,
\end{eqnarray}
CDW instability occurs. The above instability condition
 is exact up to second-order
in perturbation theory.
A more detailed and rigorous derivation of the dynamic Peierls instability
condition is given in Ref. \onlinecite{sdys}.

\section{}
We will show analytically
 that $\chi_2 (\vec{Q},0)/\chi_3 (\vec{Q},0) =3$
at 0 K. Understanding the susceptibilities is complicated
because
the eigenstates [corresponding to the eigenvalues $\lambda_{1,2}^{\vec{k}}$]
are a linear combination of the states $\psi_{k, x^2-y^2}$ and
$\psi_{k, 3 z^2-r^2}$ with coefficients that are a function of
the wavevector $\vec{k}$.
More precisely, the eigenvectors for
 $\lambda_{1,2}^{\vec{k}}$
are given by
$( c^{1\dagger}_{\vec{k}}, c^{2\dagger}_{\vec{k}}) =
(b^{\dagger}_{1 \vec{k}}, b^{\dagger}_{2 \vec{k}}) \cdot {\bf M}$, where
 ${\bf M}$ is the diagonalizing matrix for
the kinetic matrix ${\bf T}$ with
${\bf M}_{1,1}=\sin(\theta_{\vec{k}}/2)$,
${\bf M}_{2,2}=-\sin(\theta_{\vec{k}}/2)$, and
${\bf M}_{1,2}=\cos(\theta_{\vec{k}}/2)$. \\
Now, from the the kinetic matrix ${\bf T}$, we get
 \begin{eqnarray}
\cos(\theta_{\vec{p}}) =
\frac{0.5  [ \cos p_x + \cos p_y]}{\sqrt{\cos^2 p_x
+ \cos^2 p_y  - \cos p_x  \cos p_y } } ,
 \end{eqnarray}
and
 \begin{eqnarray}
\sin(\theta_{\vec{p}})=
\frac{0.5 \sqrt{3}  [ \cos p_x - \cos p_y]}{\sqrt{\cos^2 p_x
+ \cos^2 p_y  - \cos p_x  \cos p_y } } .
 \end{eqnarray}
In the expressions for $\chi_{2,3} (\vec{Q},0)$ given below
\begin{eqnarray}
 \!\!\!\!\!
  \chi_{2} (\vec{Q}, 0)=
 \sum_{\vec{k}, \alpha, \beta}
&&
 \!\!\!\!\!
\left [
 \frac
{
\langle c^{\alpha \dagger}_{\vec{k}} c^{\alpha}_{\vec{k}}\rangle
-
\langle c^{\beta\dagger}_{\vec{k}+\vec{Q}} c^{\beta}_{\vec{k}+\vec{Q}}\rangle
}
{ \lambda^{\vec{k}}_{\alpha} - \lambda^{\vec{k}+\vec{Q}}_{\beta}}
\right ] \times
\nonumber \\
&&
 \!\!\!\!\!\!\!\!\!
\sin^2 \left [
\frac{\theta_{\vec{k}+\vec{Q}} +\theta_{\vec{k}}
+ (\alpha+\beta)\pi
}{2}
\right ] ,
\label{chi-phi2}
 \end{eqnarray}
and
\begin{eqnarray}
 \!\!\!\!\!
  \chi_{3} (\vec{Q}, 0)=
 \sum_{\vec{k}, \alpha, \beta}
&&
 \!\!\!\!\!
\left [
 \frac
{
\langle c^{\alpha \dagger}_{\vec{k}} c^{\alpha}_{\vec{k}}\rangle
-
\langle c^{\beta\dagger}_{\vec{k}+\vec{Q}} c^{\beta}_{\vec{k}+\vec{Q}}\rangle
}
{ \lambda^{\vec{k}}_{\alpha} - \lambda^{\vec{k}+\vec{Q}}_{\beta}}
\right ] \times
\nonumber \\
&&
 \!\!\!\!\!\!\!\!\!
\cos^2 \left [
\frac{\theta_{\vec{k}+\vec{Q}} +\theta_{\vec{k}}
+ (\alpha+\beta)\pi
}{2}
\right ] ,
\label{chi-phi3}
 \end{eqnarray}
because $ \lambda^{\vec{k}+\vec{Q}}_{2} = -\lambda^{\vec{k}}_{1}$
and since on the Fermi surface (FS) $\lambda^{\vec{k}}_{1} =0$,
the following term diverges
\begin{eqnarray}
\left [
 \frac
{
\langle c^{1 \dagger}_{\vec{k}} c^{1}_{\vec{k}}\rangle
-
\langle c^{2\dagger}_{\vec{k}+\vec{Q}} c^{2}_{\vec{k}+\vec{Q}}\rangle
}
{ \lambda^{\vec{k}}_{1} - \lambda^{\vec{k}+\vec{Q}}_{2}}
\right ] .
 \end{eqnarray}
Furthermore, because $ \lambda^{\vec{k}+\vec{Q}}_{1} = -\lambda^{\vec{k}}_{2}$
and since  $\lambda^{\vec{k}}_{2} =0$ on the FS,
the following term also diverges
\begin{eqnarray}
\left [
 \frac
{
\langle c^{2 \dagger}_{\vec{k}} c^{2}_{\vec{k}}\rangle
-
\langle c^{1\dagger}_{\vec{k}+\vec{Q}} c^{1}_{\vec{k}+\vec{Q}}\rangle
}
{ \lambda^{\vec{k}}_{2} - \lambda^{\vec{k}+\vec{Q}}_{1}}
\right ] .
 \end{eqnarray}
Then
\begin{eqnarray}
 \chi_2 (\vec{Q},0)/\chi_3 (\vec{Q},0) && =
\frac{\cos^2 \left [
\frac{\theta_{\vec{k}+\vec{Q}} +\theta_{\vec{k}}
}{2}
\right ]_{FS}}
{\sin^2 \left [
\frac{\theta_{\vec{k}+\vec{Q}} +\theta_{\vec{k}}
}{2}
\right ]_{FS}}
\nonumber \\
&&
=
  \frac{\sin^2(\theta_{\vec{k}})_{FS}}{\cos^2(\theta_{\vec{k}})_{FS}} =3 ,
\label{chi2chi3}
 \end{eqnarray}
where use has been made of the fact that the FS is flat and
one-dimensional like and that on the FS either 
$k_x = \pm \pi/2$ or $k_y = \pm \pi/2$.\\
\\

\section{MEAN-FIELD CDW ANALYSIS}
Assuming that the total wavefunction of the system
 is separable into a phononic part and an
electronic part,
 after averaging the Hamiltonian over the phononic coordinates,
we get the following effective Hamiltonian
(with details given in Ref. \onlinecite{sudhakar}):
 \begin{eqnarray}
\bar{H}= &&
\!\!\!\!\!\!
\sum _{\vec{p}}{\bf B}^{\dagger}_{\vec{p}}\cdot{\bf T}
\cdot {\bf B}_{\vec{p}}
\nonumber \\
&&
  -
 2 g^2 \omega_0 \sum_{ j }
\left [ (b^{\dagger}_{1j} b_{2j} +
b^{\dagger}_{2j} b_{1j})
\langle b^{\dagger}_{1j} b_{2j} +
b^{\dagger}_{2j} b_{1j} \rangle \right .
\nonumber \\
&&
~~~~~~~~~~~~~~
 +
\left . (b^{\dagger}_{1j} b_{1j} -
b^{\dagger}_{2j} b_{2j})
\langle
b^{\dagger}_{1j} b_{1j} -
b^{\dagger}_{2j} b_{2j}
 \rangle \right ]
 \nonumber \\
  &&
\!\!\!\!\!\!\!\!\!\!\!\!
  +
  g^2 \omega_0 \sum_{ j }
 \langle
b^{\dagger}_{1j} b_{2j} +
b^{\dagger}_{2j} b_{1j}
 \rangle ^2
 +
\langle
b^{\dagger}_{1j} b_{1j} -
b^{\dagger}_{2j} b_{2j}
 \rangle ^2 ,
\label{2dodw}
 \end{eqnarray}
where $<..>$ implies averaging over the relevant
 coordinates which here are electronic.

\begin{figure}[t]
\includegraphics[width=3.0in]{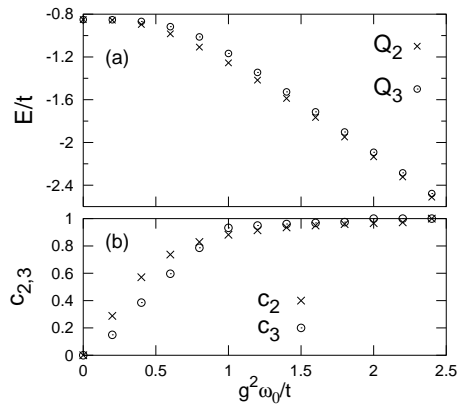}
\noindent\caption[]
{(a) Dependence of dimensionless ground state energy per site ($ E/t $)
 on dimensionless polaronic energy ($g^2 \omega_0 / t$)
for cooperative $Q_2$ and $Q_3$ modes;
(b) variation of coefficients $c_{2,3}$ of
ODW order parameters for $Q_2$ and $Q_3$ distortions as a function
of $g^2 \omega_0 / t$.}
\label{fig3}
\end{figure}

Based on the arguments that
wavevector $\vec{Q}$ determines the orbital ordering in
two-dimensions (as discussed in Sec. III), we compute the
ground state energy using mean-field
when only either $Q_2$ mode or $Q_3$ mode gets excited
cooperatively in the system. The order parameters are given by
$\langle b^{\dagger}_{1j} b_{2j} +
b^{\dagger}_{2j} b_{1j} \rangle =
 c_2 \cos (\vec{Q} \cdot \vec{R_j})$  and
$\langle b^{\dagger}_{1j} b_{1j} -
b^{\dagger}_{2j} b_{2j} \rangle  =
c_3 \cos (\vec{Q} \cdot \vec{R_j})$ with $-1 \le c_{2,3} \le 1$
and $\vec R_j$ being the position vector. Here it should be
pointed out that the order parameter
$\langle b^{\dagger}_{1j} b_{2j} +
b^{\dagger}_{2j} b_{1j} \rangle$
corresponds to the density difference of electrons in the
two orbitals
 $\psi_X \equiv (\psi_{x^2-y^2} - \psi_{3z^2 -r^2})/\sqrt{2}$
 and $ \psi_Y \equiv -(\psi_{x^2-y^2} + \psi_{3z^2 -r^2})/\sqrt{2}$ (as
described in 
Ref. \onlinecite{allen}).

The unit cell needed to
compute the ground state energy
consists of two adjacent sites with the Brillouin zone being given by
$-\pi ~\le (k_x + k_y) ~\le \pi $
and $-\pi ~\le (k_x - k_y) ~\le \pi $. We
diagonalize
a $4 \times 4$ matrix at each momentum and integrate the lowest
two eigenenergies over the Brillouin zone to obtain the ground state
energy.
The results of
our calculations are shown in Fig. \ref{fig3}. From Fig. \ref{fig3}(a)
we see that the ground state energy corresponds to the $Q_2$ mode
with the difference in energy between the $Q_2$ only state and the $Q_3$
only state peaking at intermediate values of the dimensionless polaronic energy
($g^2 \omega_0 / t$).
For zero values and infinite values of
the polaronic energy both modes yield the same energy
because zero value implies no phononic coupling effect while infinite value
corresponds to localized polarons.
Thus for
large values of the polaronic energy,
the ground  state energy is only slightly smaller than the polaronic energy.
Furthermore, from Fig. \ref{fig3}(b) we also see that, 
as the polaronic energy increases,
the values of $c_{2,3}$ increase and become unity around
 $g^2\omega_0 / t \sim 2$ implying that for the $Q_3$ ($Q_2$) mode
$\psi_{x^2-y^2}$ ($\psi_X$) orbital is occupied
fully at one site with the $\psi_{3z^2 -r^2}$ ($\psi_Y$)
orbital being fully occupied at the adjacent sites.

\end{document}